\documentclass[letterpaper]{article} 
\usepackage{aaai2026}  
\usepackage{times}  
\usepackage{helvet}  
\usepackage{courier}  
\usepackage[hyphens]{url}  
\usepackage{graphicx} 
\usepackage{natbib}  
\usepackage{caption} 
\usepackage{algorithm}
\usepackage{algorithmic}

\usepackage{newfloat}
\usepackage{listings}
\DeclareCaptionStyle{ruled}{labelfont=normalfont,labelsep=colon,strut=off} 
\floatstyle{ruled}
\newfloat{listing}{tb}{lst}{}
\floatname{listing}{Listing}
\title{Yours or Mine? Overwriting Attacks Against Neural Audio Watermarking}
\author{
    Lingfeng Yao\equalcontrib\textsuperscript{\rm 1}, Chenpei Huang\equalcontrib\textsuperscript{\rm 1}, Shengyao Wang\textsuperscript{\rm 2}, Junpei Xue\textsuperscript{\rm 2}, \\Hanqing Guo\textsuperscript{\rm 3}, Jiang Liu\textsuperscript{\rm 2}, Phone Lin\textsuperscript{\rm 4}, Tomoaki Ohtsuki\textsuperscript{\rm 5}, Miao Pan\textsuperscript{\rm 1} \\
}
\affiliations{
    \textsuperscript{\rm 1} University of Houston\\
    \textsuperscript{\rm 2} Waseda University\\
    \textsuperscript{\rm 3} University of Hawaii at Mānoa,\\
    \textsuperscript{\rm 4} National Taiwan University,\\
    \textsuperscript{\rm 5} Keio University
    \\lyao12@uh.edu, chuang30@uh.edu, swan@ruri.waseda.jp, akane123@akane.waseda.jp, \\guohanqi@hawaii.edu, jiang@waseda.jp, plin@csie.ntu.edu.tw, ohtsuki@keio.jp, mpan2@uh.edu
}

\usepackage{bibentry}
\usepackage{xcolor}
\usepackage{booktabs}
\usepackage{amsmath} 
\usepackage{amsfonts} 
\usepackage{dsfont}
\usepackage{subcaption}

\begin{document}
\maketitle
\begin{abstract}
As generative audio models are rapidly evolving, AI-generated audios increasingly raise concerns about copyright infringement and misinformation spread. Audio watermarking, as a proactive defense, can embed secret messages into audio for copyright protection and source verification. However, current neural audio watermarking methods focus primarily on the imperceptibility and robustness of watermarking, while ignoring its vulnerability to security attacks. In this paper, we develop a simple yet powerful attack: the overwriting attack that overwrites the legitimate audio watermark with a forged one and makes the original legitimate watermark undetectable. Based on the audio watermarking information that the adversary has, we propose three categories of overwriting attacks, i.e., white-box, gray-box, and black-box attacks. We also thoroughly evaluate the proposed attacks on state-of-the-art neural audio watermarking methods. Experimental results demonstrate that the proposed overwriting attacks can effectively compromise existing watermarking schemes across various settings and achieve a nearly 100\% attack success rate. The practicality and effectiveness of the proposed overwriting attacks expose security flaws in existing neural audio watermarking systems, underscoring the need to enhance security in future audio watermarking designs.
\end{abstract}

\section{Introduction}
With the rapid development of generative audio models, artificially producing highly realistic speeches is becoming accessible. This progress also introduces new social risks. For example, attackers can exploit these models to clone or impersonate a person's voice, enabling voice fraud or copyright infringement. These risks arise from a key limitation: human listeners struggle to tell AI-generated audio from real speech audio. To address this problem, audio watermarking has emerged as a proactive defense mechanism. By embedding imperceptible digital signatures into audio signals, watermarking enables future verification of copyright ownership or identification of the audio's source.

To improve usability, existing neural audio watermarking methods primarily focus on two properties: robustness and imperceptibility. Robustness ensures that the watermark remains detectable after common signal processing operations such as compression. Imperceptibility guarantees that watermark embedding does not degrade perceptual audio quality. These objectives have driven recent progress in watermarking research. AudioSeal~\cite{san2024proactive} is based on the Encodec~\cite{defossez2022high} architecture, embedding redundant watermark messages in the embedding layer to improve robustness, while introducing novel perceptual loss to preserve audio quality. Timbre~\cite{liu2024detecting} embeds the watermark in the frequency domain and incorporates a distortion layer during training to simulate real-world perturbations, thereby maintaining robustness. WavMark~\cite{chen2023wavmark} adopts invertible neural networks~\cite{dinh2014nice} to embed imperceptible watermarks and use pattern bits to improve robustness against distortions.

Although the robustness of audio watermarking has been extensively studied~\cite{wen2025sok, shallowwatermark, ozer2025comprehensive}, its security aspects remain underexplored. Robustness refers to the ability to withstand unintentional perturbations, whereas security concerns its resilience against intentional manipulation by adversaries~\cite{hartung2002multimedia, furon2003asymmetric, li2021talk}. \citet{san2024proactive} was the first to highlight potential security threats in neural audio watermarking. It showed that adversaries, with access to the watermark detector, can launch two adversarial attacks: removal attacks, which make the watermark undetectable, and forgery attacks, which falsely embed a watermark into clean audio. \citet{liu2024audiomarkbench} extended these attack paradigms and systematically evaluated the vulnerabilities of neural audio watermarking methods under such threats. Furthermore,  \citet{liu2024detecting} discussed overwriting attacks, demonstrating that an attacker with access to the watermark embedder can insert a new watermark that effectively overwrites the original one. However, their attack relies on a white-box assumption, where the adversary has full knowledge of the watermarking framework, and the evaluation of the overwriting attack lacks thorough analysis.

In this work, we present the first systematic study of watermark overwriting attacks, a previously underexplored but practically powerful threat. Unlike removal or forgery attacks, overwriting attacks embed a new watermark to replace the original legitimate one and thus hijack the ownership of the target audio. We perform comprehensive evaluations of three state-of-the-art neural audio watermarking methods~\cite{san2024proactive,liu2024detecting,chen2023wavmark} under the proposed white-box, gray-box, and black-box overwriting attacks. In all threat settings, we achieve a nearly 100\% attack success rate in terms of overwriting the original watermark. These findings expose widespread vulnerabilities in existing neural audio watermarking systems and underscore the need to consider security as a primary design objective, alongside robustness and imperceptibility. Our main contributions are summarized as follows.

\begin{itemize}
    \item We present the first systematic study of overwriting attacks, a powerful yet underexplored security threat in neural audio watermarking. Our work thoroughly analyzes the mechanisms of such attacks and their effects across various threat models.
    \item Based on the audio watermarking information the adversary has, we propose three-level overwriting attacks, i.e., white-box, gray-box, and black-box attacks, and develop the corresponding attack procedures.
    \item Through extensive experiments on three state-of-the-art watermarking methods, we find that the proposed overwriting attacks achieve nearly 100\% attack success rates. These analyses and experimental results validate that the overwriting attack is a fundamental security challenge to existing neural audio watermarking designs.
\end{itemize}

\section{Background and Related Work}\label{background_and_relatedwork}
\subsection{Principles of Audio Watermarking }
Audio watermarking is a technique that embeds information into audio signals without significantly degrading their perceptual quality. A typical audio watermarking system consists of two key components: a \textbf{watermark embedder}, which encodes the information into the audio, and a \textbf{watermark detector}, which detects and recovers the embedded message. The watermark is typically a fixed-length binary sequence and can carry various types of information depending on the applications. For example, when the watermark encodes copyright metadata, it can support copyright declarations and infringement tracking~\cite{liu2024detecting}; when it includes source-related tags (e.g., indicators of AI-generated content), it enables source verification~\cite{san2024proactive}; and when it contains a hash of the audio, it can be used for integrity verification and tampering detection~\cite{yao2025speechverifier}.

As a proactive protection mechanism, audio watermarking systems typically require the following three properties~\cite{hartung2002multimedia}: \textbf{robustness}, \textbf{imperceptibility}, and \textbf{security}. Robustness refers to the system's ability to retain the watermark after undergoing common audio processing or transmission (e.g., resampling, compression, or reverberation). Imperceptibility requires that the watermark embedding process introduces no perceptible distortion to audio. Security emphasizes resilience against intentional attacks, ensuring the watermark's integrity even when adversaries attempt to manipulate it.

\subsection{Neural Audio Watermarking Methods}
Audio watermarking has evolved from traditional signal processing techniques into deep learning-based systems, achieving significant progress. Traditional methods, such as least significant bit (LSB)~\cite{cvejic2004increasing}, echo hiding~\cite{gruhl1996echo}, spread spectrum~\cite{cox1997secure}, patchwork~\cite{yeo2003modified}, and quantization index modulation (QIM)~\cite{chen2002quantization}, typically rely on expert knowledge and fixed rules. They are hard to implement and struggle with robustness against complex distortions.

Recent developments in deep learning have introduced a new paradigm for audio watermarking. End-to-end neural watermarking models enable joint optimization of the embedding and detection processes. By leveraging carefully designed loss functions, these models are able to simultaneously enhance robustness and maintain imperceptibility.

Recent neural audio watermarking methods can be broadly categorized into three classes based on their embedding strategies:
\begin{itemize}
    \item \textbf{Encoder-decoder-based approaches}, such as AudioSeal~\cite{san2024proactive}, XattnMark~\cite{liu2025xattnmark}, and SilentCipher~\cite{singh2024silentcipher}, embed watermarks in the high-dimensional latent space learned by neural networks;
    \item \textbf{Frequency-domain-based approaches}, such as Timbre~\cite{liu2024detecting} and DeAR~\cite{liu2023dear}, embed watermarks into the frequency spectrum of audio;
    \item \textbf{Invertible neural network (INN)-based approaches}, such as WavMark~\cite{chen2023wavmark} and IDEAW~\cite{li2024ideaw}, model the embedding and detection processes as reversible transformations, enabling high-fidelity and accurate watermark recovery.
\end{itemize}

In this work, we select three representative systems, AudioSeal, Timbre, and WavMark, from the respective categories. We systematically evaluate their vulnerability to overwriting attacks, a practical and previously underexplored attack in neural audio watermarking.

\subsection{Security Challenges and Existing Attacks}
In audio watermarking systems, security and robustness are two closely related yet fundamentally different properties. Robustness refers to the ability to withstand benign processing operations that are not intended to affect the watermark. In contrast, security concerns its resilience against intentional attacks, where adversaries aim to remove, forge, or manipulate the watermark, and may even possess partial or full access to the watermarking system.

Most existing neural audio watermarking methods prioritize robustness and imperceptibility, while largely overlooking security. Traditional audio watermarking methods~\cite{furon2003asymmetric} often leverage the secret key to control the embedding location and detection process, thus preventing unauthorized manipulation. In contrast, neural watermarking systems typically lack such explicit key-based security mechanisms, but rely on the assumption that the secret of the model weights. However, this assumption is fragile in modern research environments, where open-sourcing and reverse engineering are common practices. According to \textbf{Kerckhoffs's principle}~\cite{kerckhoffs1883cryptographie}, a secure system should remain secure even if everything about the system is public except the secret key. Therefore, relying on ``security through obscurity'' is inadequate for neural watermarking systems.

Recent studies have begun to explore the security vulnerabilities of neural watermarking. \citet{san2024proactive} first demonstrated that adversaries with access to the watermark detector could launch two types of adversarial attacks: removal attacks, which make the watermark undetectable, and forgery attacks, which falsely embed a watermark into clean audio. \citet{liu2024audiomarkbench} extended this line of work with a more systematic evaluation of neural audio watermarking methods under adversarial attacks. 

However, \textbf{overwriting attacks}, where adversaries embed a new watermark into an already watermarked audio to override the original ownership, remain largely underexplored. This attack poses a realistic and severe threat to ownership verification. Although \citet{liu2024detecting} pointed out overwriting attacks in white-box scenarios, their discussion was limited to insider threats and did not provide a comprehensive evaluation across different adversarial settings. In this work, we address this gap by systematically investigating overwriting attacks across various threat models, detailing their implementations and evaluating their effects on existing neural audio watermarking systems.

\section{System and Threat Model}
\subsection{System Model}
Let $x$ denote a clean audio and $m \in \{0,1\}^L$ represent an $L$-bit binary message intended for embedding. A neural audio watermarking system comprises two key components: an embedder $\mathcal{E}$ and a detector $\mathcal{D}$. The embedder embeds the message into the audio, generating a watermarked signal $x_w = \mathcal{E}(x, m)$. The detector recovers the message $\hat{m}$ from the watermarked audio as $\hat{m} = \mathcal{D}(x_w)$. Ideally, the recovered message matches exactly the embedded message, i.e., $\hat{m} = m$. In practice, a watermarking system must satisfy the following three properties.

\paragraph{Imperceptibility.} The watermark embedding process is required to avoid perceptible distortions, which is formally expressed as: $d(x, x_w) \leq \epsilon$, where $d(\cdot,\cdot)$ is a perceptual distance metric and $\epsilon$ is an auditory tolerance threshold.

\paragraph{Robustness.} The embedded watermark is expected to survive common, unintentional audio transformations $\mathcal{T(\cdot)}$, such as resampling and compression, i.e., $\mathcal{D}(\mathcal{T}(x_w)) = m$.

\paragraph{Security.} The watermark is expected to resist adversarial or intentional manipulations $\delta(\cdot)$, which aim to remove, alter, or overwrite the legitimate watermark, satisfying $\mathcal{D}(\delta(x_w)) = m$.

\noindent Previous work focuses on imperceptibility and robustness, while the security property remains underexplored. To bridge this gap, this work systematically investigates an intuitive but practically powerful attack: overwriting attacks on neural audio watermarking systems.    

\subsection{Threat Model}
\begin{figure}[!t]
\centering
\includegraphics[width=1.0\columnwidth]{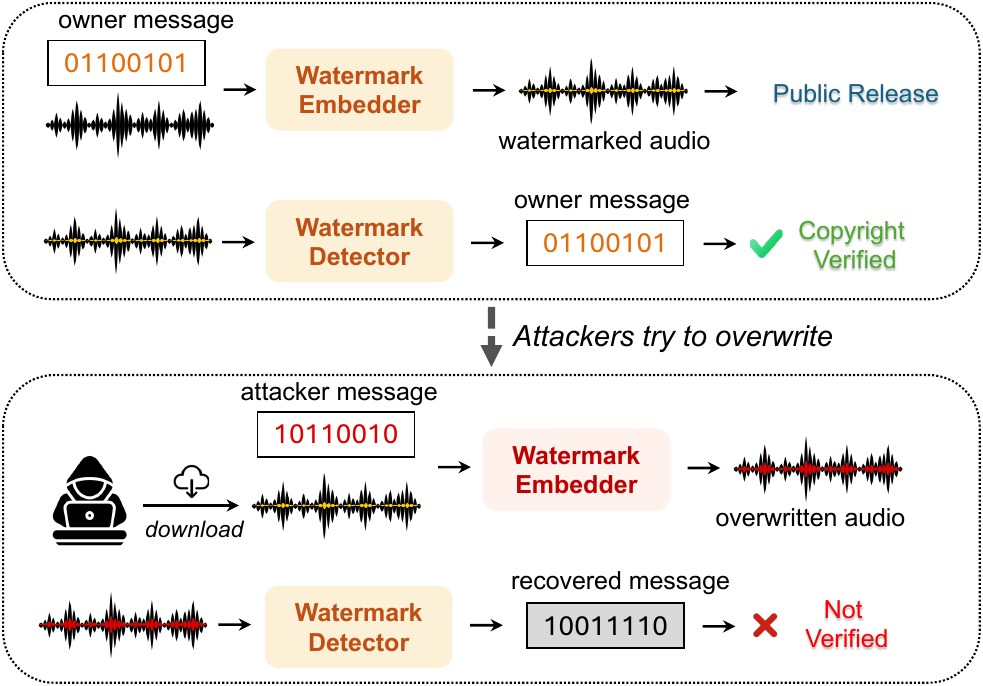}
\caption{Overview of the proposed audio watermark overwriting attack. An adversary injects a new forged watermark into an already watermarked audio, erasing the original legitimate watermark. Thus, legitimate ownership cannot be verified, and the adversary can falsely claim the copyright.}
\label{attack_process}
\end{figure}

The core threat studied here is the overwriting attack, as shown in Figure~\ref{attack_process}. The adversary replaces the original watermark with their own, thus hijacking the audio ownership.

\paragraph{Adversarial Goal.}
To understand the adversarial goal, we first outline the standard copyright verification protocol: a legitimate owner embeds a secret message $m_{owner}$ into audio $x$, creating a watermarked version $x_w = \mathcal{E}(x,m_{owner})$, which is then publicly distributed. If ownership disputes arise, the owner reveals $m_{owner}$. Then, the arbiter verifies the ownership by checking whether the detector recovers the matched message, i.e., $\mathcal{D}(x_w)=m_{owner}$.

\noindent The adversarial goal is to break this protocol. Given a publicly available watermarked audio $x_w$, the adversary uses another embedder $\mathcal{E}'$ to embed a new message $m_{adv}'$, and thus generates a forged audio $x_w' = \mathcal{E}'(x_w, m_{adv}')$. A successful attack satisfies the following conditions: (i) The original message is no longer recoverable: $\mathcal{D}(x_w') \neq m_{owner}$, (ii) the adversary's detector accurately extracts their own message: $\mathcal{D}'(x_w') = m_{adv}'$, and (iii) the forged audio remains imperceptible from the original one: $d(x_w', x_w) \leq \epsilon$.

\paragraph{Attacker Capabilities.}
Based on the adversary's knowledge of legitimate audio watermarking systems, we propose the following overwriting attacks correspondingly.

\begin{itemize}
    \item \textbf{White-box:} The adversary has full access to the original watermarking embedder $\mathcal{E}$, representing insider threats or attacks on fully open-sourced watermarking models.
    \item \textbf{Gray-box:}  The adversary has partial knowledge. They know the general architecture of the watermarking models (e.g., SOTA watermarking designs), but lack the knowledge of model weights and training details. To implement such an attack, the adversary must train a surrogate model $\mathcal{E}'$ to replicate and further replace the functionality of the target watermark.
    \item \textbf{Black-box:} The adversary has no knowledge of the model's architecture or its weights. The black-box overwriting attacks can be classified into two subcategories.
    \begin{itemize}
        \item \textbf{Query-based:} The adversary has limited API access to the original legitimate detector $\mathcal{D}$ output. They can make a few queries to infer the specific watermarking system and then train a surrogate embedder.
        \item \textbf{Zero-query:} The adversary has no query access to the detector $\mathcal{D}$. They apply a set of public watermarking models or retrained surrogate models to the target watermarked audio speech in a brute-force manner.
    \end{itemize}
\end{itemize}

\section{Overwriting Attack Designs}
In this section, we present the proposed overwriting attack designs in detail. We aim to embed an adversarial message $m_{adv}'$ into an already legitimately watermarked audio $x_w$ and then generate an overwritten audio $x_w'$. $m_{adv}'$ is an arbitrary binary sequence selected by the attacker to represent forged ownership or other identifying information.

\subsection{White-box Attack}
In the white-box setting, the attacker has full knowledge and access to the original legitimate watermarking embedder. Thus, the overwriting attack is straightforward:
\begin{equation}
    x_w' = \mathcal{E}(x_w,m_{adv}').
\end{equation}

\subsection{Gray-box Attack}
In the gray-box setting, the adversary lacks access to the weights, training data, and precise training details (e.g., loss functions) of the original watermarking model. Therefore, adversaries must train a surrogate watermarking system ($\mathcal{E}',\mathcal{D}'$). We propose a general watermark training framework to achieve this, which optimizes a joint loss to balance message embedding accuracy and audio imperceptibility.
\begin{equation}
\mathcal{L}_{\text{total}} = \lambda_w \cdot \mathcal{L}_w + \lambda_t \cdot \mathcal{L}_{\text{recon}_t} + \lambda_f \cdot \mathcal{L}_{\text{recon}_f} + \lambda_{\text{adv}} \cdot \mathcal{L}_{\text{adv}},
\end{equation}
where $\lambda_w$, $\lambda_t$, $\lambda_f$, and $\lambda_{\text{adv}}$ are hyperparameters to balance these terms. The individual loss components are as follows.
\begin{itemize}
    \item \textbf{Watermark Recovery Loss ($\mathcal{L}_{w}$)}: To ensure accurate embedding and detection, we apply binary cross-entropy (BCE) loss between the input message $m$ and the detected message from the surrogate system:
    \begin{equation}
    \mathcal{L}_{\text{w}} = \mathrm{BCE}(m, \mathcal{D}'(\mathcal{E}'(x, m))).
    \end{equation}

    \item \textbf{Time-domain Reconstruction Loss ($\mathcal{L}_{\text{recon}_t}$)}: To minimize audible distortions, we employ mean squared error (MSE) between the clean and watermarked audio signals:
    \begin{equation}
    \mathcal{L}_{\text{recon}_t} = \mathrm{MSE}(x, \mathcal{E}'(x, m)).
    \end{equation}

    \item \textbf{Frequency-domain Reconstruction Loss ($\mathcal{L}_{\text{recon}_f}$)}: To further reduce perceptual differences in the frequency domain, we introduce a multi-resolution short-time Fourier transform (STFT) loss~\cite{yamamoto2020parallel}. For each resolution $m$, the loss includes a spectral convergence and a log-magnitude term:
    \begin{align}
    \mathcal{L}_{\text{sc}}^{(m)} &= \frac{\| S_m(x) - S_m(\mathcal{E}'(x, m)) \|_F}{\| S_m(x) \|_F}, \\
    \mathcal{L}_{\text{mag}}^{(m)} &= \frac{1}{N} \| \mathrm{\log}(S_m(x)) - \mathrm{\log}(S_m(\mathcal{E}'(x, m))) \|_1,
    \end{align}
    where $S_m(\cdot)$ denotes the STFT operation at resolution $m$, and $N$ is the number of spectrogram elements. The aggregated frequency-domain loss is computed as:
    \begin{equation}
    \mathcal{L}_{\text{recon}_f} = \frac{1}{M} \sum_{m=1}^M \left( \mathcal{L}_{\text{sc}}^{(m)} + \mathcal{L}_{\text{mag}}^{(m)} \right).
    \end{equation}
    
    \item \textbf{Adversarial Loss ($\mathcal{L}_{\text{adv}}$)}: To further enhance perceptual quality, we employ adversarial training. A discriminator $D$ is trained to distinguish $x$ from $\mathcal{E}'(x, m)$, while the surrogate embedder aims to generate watermarked audio indistinguishable from the original one, i.e.,
    \begin{equation}
    \mathcal{L}_{\text{d}} = -\mathrm{\log}(\sigma(D(x))) - \mathrm{\log}(1 - \sigma(D(\mathcal{E}'(x, m)))),
    \end{equation}
    where $\sigma(\cdot)$ denotes the sigmoid function. The adversarial loss for the embedder is:
    \begin{equation}
    \mathcal{L}_{\text{adv}} = -\mathrm{\log}(\sigma(D(\mathcal{E}'(x, m)))).
    \end{equation}
    \end{itemize}

\noindent After training, the surrogate embedder $\mathcal{E}'$ is ready for overwriting attacks as defined in the white-box scenario.

\subsection{Black-box Attack}
In the black-box setting, adversaries have no knowledge of the original model architecture or its parameters. Attacking strategies vary according to query accessibility:
\begin{itemize}
    \item \textbf{Zero-query Attack:} Without any query access, adversaries collect or reproduce a set of public watermarking models $\mathcal{E}_i'$, and then apply them as much as possible in a brute-force manner to overwrite the original watermark.
    \begin{equation}
    x_w^{(N)} = \left( \mathcal{E}_N \circ \mathcal{E}_{N-1} \circ \cdots \circ \mathcal{E}_1 \right)(x_w, m_{adv}^{'}),
    \end{equation}
    where $\mathcal{E}_i$ denotes the $i$-th surrogate watermark embedder collected or trained by adversaries, $x_w^{(N)}$ is the resulting audio after sequentially applying $N$ surrogate embedders, and $\circ$ denotes the function composition operation, which indicates the sequential application from the innermost to the outermost embedder.
    
    \item \textbf{Query-based Attack:} With limited query access to the original detector $\mathcal{D}$, a more efficient strategy is possible.
    \begin{itemize}
        \item Partially train the candidate surrogate models for a limited number of epochs.
        \item Use these undertrained models to embed new messages into $x_w$;
        \item Query the original detector $\mathcal{D}$ to evaluate whether the original watermark has been tampered;
        \item Identify the most effective candidate and refine training until it can reliably perform overwriting attacks.
    \end{itemize}
\end{itemize}

\noindent The query-guided black-box attack significantly reduces computational costs while increasing attacking efficacy.

\section{Experiments and Analyses}
To evaluate the vulnerability of neural audio watermarking systems to overwriting attacks, we conduct extensive experiments on three representative methods: AudioSeal~\cite{san2024proactive}, Timbre~\cite{liu2024detecting}, and WavMark~\cite{chen2023wavmark}. Evaluations are performed under three threat models: white-box, gray-box, and black-box.

\subsection{Experiment Settings}
\paragraph{Datasets and Training Setup.}
We conduct experiments on two widely used speech datasets: \textbf{LibriSpeech}~\cite{panayotov2015librispeech}, a corpus of approximately 1,000 hours of English read speech, and \textbf{VoxCeleb1}~\cite{nagraniy2017voxceleb}, which contains over 150,000 utterances from 1,251 celebrities. All audio samples are resampled to 16kHz and converted to WAV format for consistency across models. All models are trained on a server equipped with 64 CPU cores and two NVIDIA A100 GPUs.

\paragraph{Metrics.}
We assess attack performance using the following metrics:
\begin{itemize}
    \item \textbf{Bit Error Rate (BER)} measures the proportion of incorrectly recovered bits. Given the embedded message $m \in \{0,1\}^L$ and the detected message $\hat{m}$:
    \begin{equation}
    \text{BER} = \frac{1}{L} \sum_{i=1}^{L} \mathds{1}\left[m_i \ne \hat{m}_i\right],
    \end{equation}

    \item \textbf{Attack Success Rate (ASR)} reflects how often the embedded message is corrupted after the overwriting attack. It is computed as the proportion of samples where the detected message differs from the original:
    \begin{equation}
    \text{ASR} = \frac{1}{N} \sum_{j=1}^{N} \mathds{1}\left[ \mathcal{D}(x_{w,j}') \ne m_j \right],
    \end{equation}
    where $x_{w,j}'$ is the $j$-th overwritten audio, $N$ is the total number of evaluated samples, and $\mathds{1}[\cdot]$ is the indicator function.
        
    \item \textbf{Signal-to-Noise Ratio (SNR)} quantifies the audio distortion introduced by overwriting. It compares the power of the original watermarked audio $x_w$ and the overwritten audio $x_w'$:
        \begin{equation}
        \text{SNR} = 10 \log_{10} \left( \frac{|x_w|_2^2}{|x_w - x_w'|_2^2} \right),
        \end{equation}
\end{itemize}

\subsection{Overwriting Attack Results}
\begin{figure}[t]
\centering
\includegraphics[width=0.6\columnwidth]{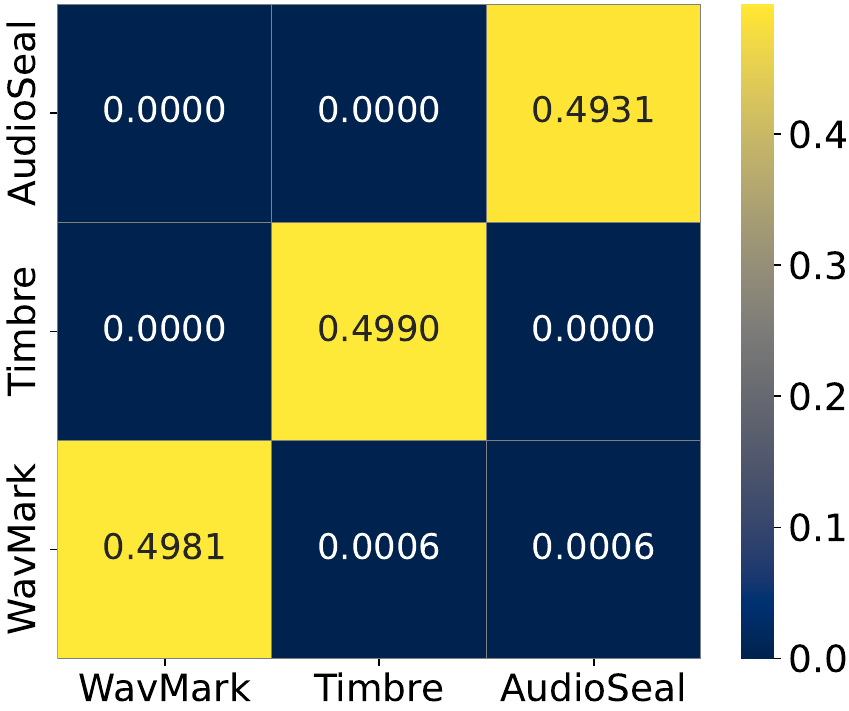}
\caption{Bit error rate (\%) of the original watermark after white-box overwriting.}
\label{whitebox_confusion_matrix}
\end{figure}

We first consider the white-box setting, where the attacker has full access to the original legitimate watermarking model. Figure~\ref{whitebox_confusion_matrix} shows the BER between the original message and the recovered message after being overwritten. The x-axis represents the watermarking method used to embed the original message, and the y-axis denotes the one used by the attacker to embed the overwriting watermark.

\paragraph{White-box Attack.}

The diagonal entries (the same method used for both original and overwriting watermarks) consistently reach a BER near 0.5, which means random guessing. This confirms that the original watermark is completely disrupted when the same watermarking method is reused for overwriting. In contrast, off-diagonal values (different methods) show very low BERs, suggesting that overwriting using a different watermarking method fails to destroy the original legitimate watermark. This is because different watermarking methods operate in distinct embedding domains and rely on method-specific decoding mechanisms.

\begin{table}[!t]
\setlength{\tabcolsep}{11pt}
\centering
\begin{tabular}{l|ccc}
\toprule
\textbf{Metric} & \textbf{Timbre} & \textbf{AudioSeal} & \textbf{WavMark} \\
\midrule
ASR & 99.80 & 100.00 & 100.00 \\
ACC & 100.00 & 100.00 & 100.00 \\
\bottomrule
\end{tabular}
\caption{White-box overwriting results. Attack success rate (ASR) of the original watermark and recovery accuracy (ACC) of the overwritten watermark (\%).}
\label{tab:whitebox_results}
\end{table}

\begin{figure*}[t]
\centering
\begin{subfigure}[b]{0.45\textwidth}
    \centering
    \includegraphics[width=\textwidth]{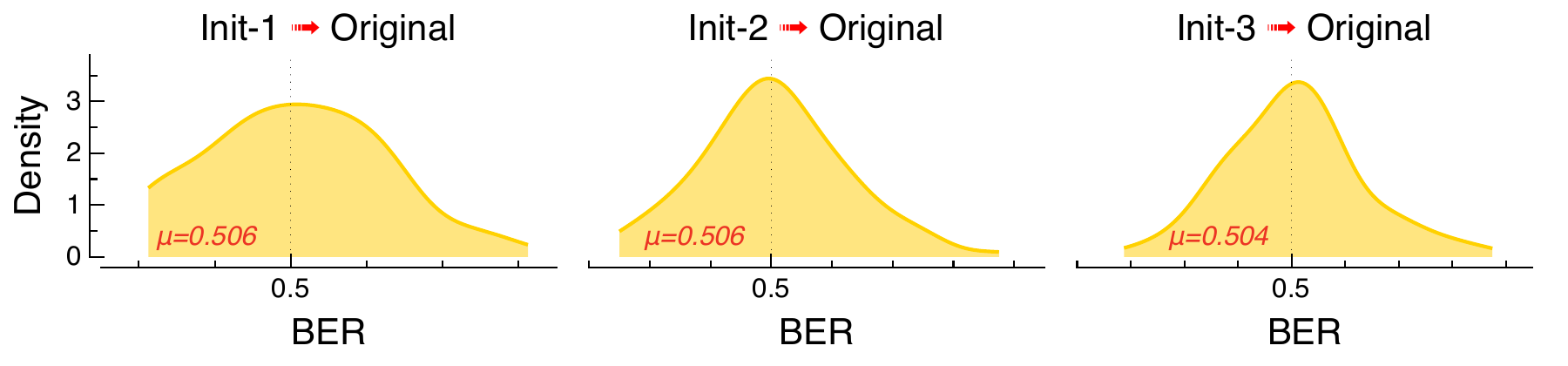}
    \caption{BER of AudioSeal (cross-training).}
    \label{graybox_crosstraining_audioseal}
\end{subfigure}
\hfill
\begin{subfigure}[b]{0.45\textwidth}
    \centering
    \includegraphics[width=\textwidth]{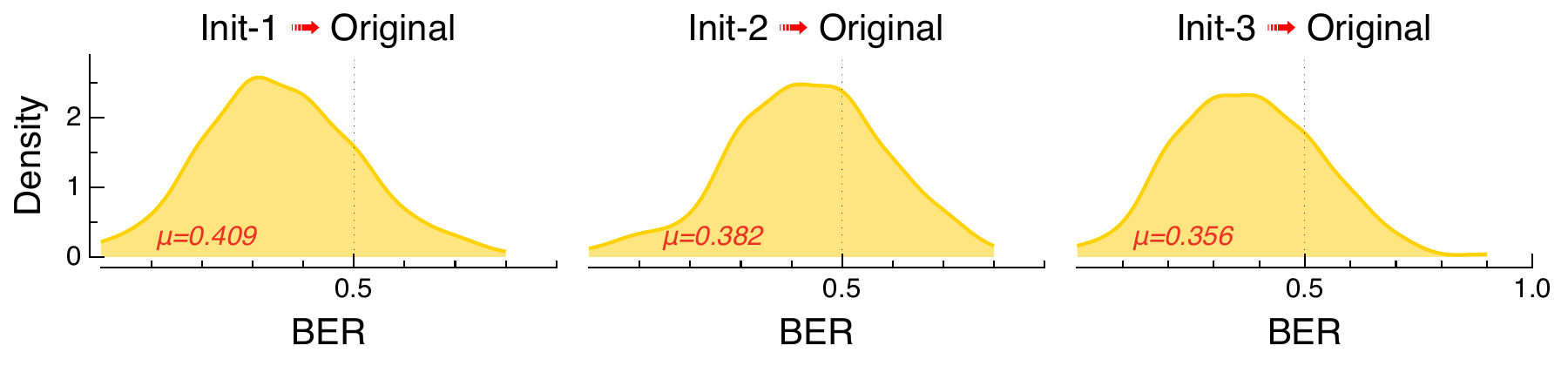}
    \caption{BER of Timbre (cross-training).}
    \label{graybox_crosstraining_timbre}
\end{subfigure}

\begin{subfigure}[b]{0.45\textwidth}
    \centering
    \includegraphics[width=\textwidth]{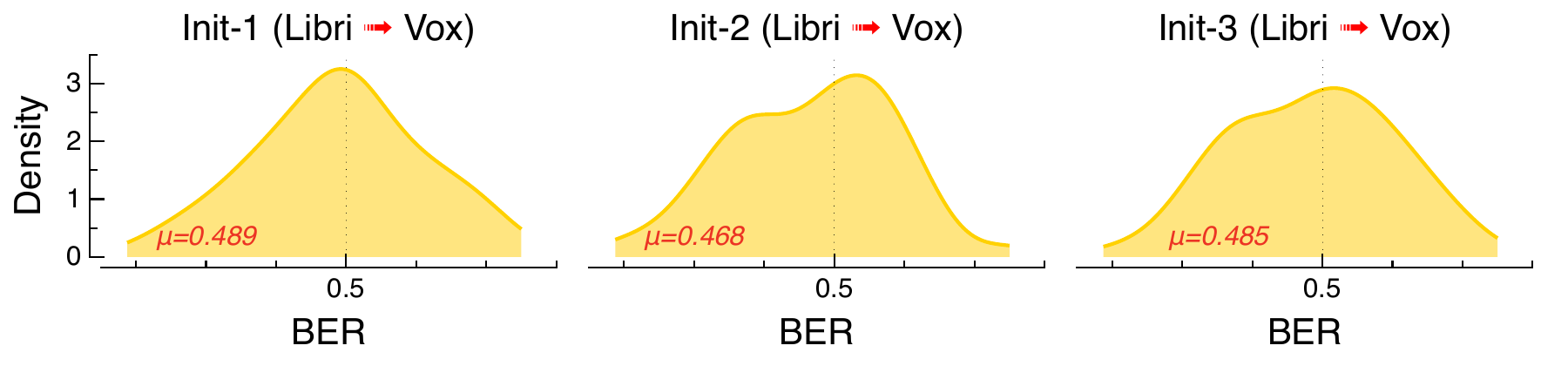}
    \caption{BER of on AudioSeal (cross-data).}
    \label{graybox_crossdata_audioseal}
\end{subfigure}
\hfill
\begin{subfigure}[b]{0.45\textwidth}
    \centering
    \includegraphics[width=\textwidth]{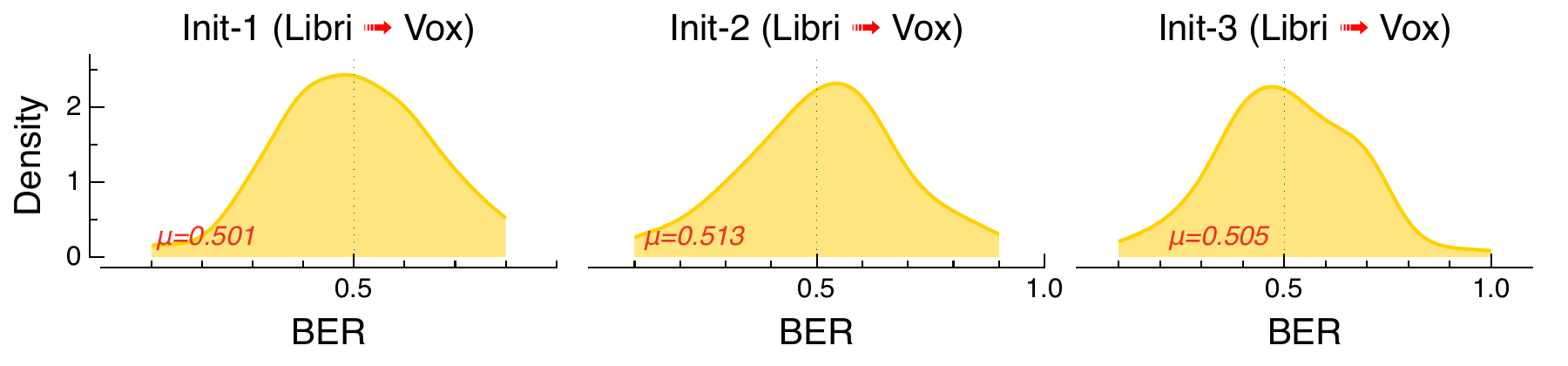}
    \caption{BER of Timbre (cross-data).}
    \label{graybox_crossdata_timbre}
\end{subfigure}
\caption{Bit error rate (\%) distributions of the original watermark for AudioSeal and Timbre under gray-box settings.}
\label{fig:graybox_2x2}
\end{figure*}

Table~\ref{tab:whitebox_results} summarizes the attack success rate (ASR) of the original watermark and watermark recovery accuracy of the overwritten watermark. All three methods achieve an ASR near 100\%, indicating that the original watermark is no longer verifiable. Meanwhile, the overwritten watermark is recovered with perfect accuracy, indicating that the attacker has successfully hijacked the audio ownership.

These results reveal a critical vulnerability: when the attackers have access to the same watermarking embedder, they can reliably overwrite the original legitimate watermark with their own, effectively hijacking the audio ownership.

\begin{table}[t]
\setlength{\tabcolsep}{14pt}  
\centering
\begin{tabular}{lccc}
\toprule
\textbf{Method} & Init-1 & Init-2 & Init-3 \\
\midrule
\multicolumn{4}{l}{\textit{ASR (Original Watermark)}} \\
Timbre     & 99.60 & 98.80 & 98.40 \\
AudioSeal  & 100.00 & 100.00 & 100.00 \\
WavMark    & 100.00 & 100.00 & 99.50 \\
\midrule
\multicolumn{4}{l}{\textit{ACC (Overwritten Watermark)}} \\
Timbre     & 100.00 & 100.00 & 100.00 \\
AudioSeal  & 99.40 & 99.00 & 99.80 \\
WavMark    & 100.00 & 100.00 & 100.00 \\
\bottomrule
\end{tabular}
\caption{Gray-box cross-training results (\%). Top: attack success rate (ASR) of the original watermark. Bottom: recovery accuracy (ACC) of the overwritten watermark.}
\label{tab:graybox_crosstraining_combined}
\end{table}

\paragraph{Gray-box Attack.} In the gray-box setting, the adversary knows the watermarking system's architecture but lacks access to its training details or training data. To assess this scenario, we construct surrogate models under two settings: (i) cross-training, where the surrogate model is trained on the same dataset (VoxCeleb1) but with a different training pipeline, and (ii) cross-data, where the surrogate is trained on an entirely different dataset (LibriSpeech). We train three surrogate models with varying random seeds, denoted as Init-1, Init-2, and Init-3, respectively. The “Init-$k$$\rightarrow$Official” notation indicates that surrogate Init-$k$ attempts to overwrite the watermark embedded by official model.

Figures~\ref{graybox_crosstraining_audioseal} and~\ref{graybox_crosstraining_timbre} present the bit error rate (BER) distributions of the original watermark under cross-training attacks. For AudioSeal, all surrogate models yield BER distributions tightly centered around 0.5 (mean $\mu$ = 0.504 - 0.506). This indicates a complete corruption of the original watermark, as the detection performance is consistent with random guessing. In contrast, attacks on Timbre yielded distributions around 0.4, suggesting that while most bits of the watermark are corrupted, some residual information still remains. We exclude WavMark from the BER distribution analyses because its decoder uses pattern bit verification: once the watermark is overwritten, the pattern check fails and the decoder outputs nothing, which makes BER impossible to define. Therefore, we report only ASR and recovery accuracy for WavMark.

The cross-training quantitative results are summarized in Table~\ref{tab:graybox_crosstraining_combined}. For all watermarking methods, the attack success rate approaches 100\%, demonstrating that surrogate models consistently invalidate the original watermark. Simultaneously, the overwritten watermark is recovered with near-perfect accuracy. This suggests that different training configurations do not significantly alter the embedding behavior to resist overwriting attacks.

\begin{table}[t]
\setlength{\tabcolsep}{8pt}
\centering
\begin{tabular}{lccc}
\toprule
\textbf{Method (Libri $\rightarrow$ Vox)} & Init-1 & Init-2 & Init-3 \\
\midrule
\multicolumn{4}{l}{\textit{ASR (Original Watermark)}} \\
Timbre     & 99.80 & 99.90 & 98.80 \\
AudioSeal  & 100.00 & 100.00 & 100.00 \\
WavMark    & 100.00 & 100.00 & 100.00 \\
\midrule
\multicolumn{4}{l}{\textit{ACC (Overwritten Watermark)}} \\
Timbre     & 99.90 & 100.00 & 99.90 \\
AudioSeal  & 97.20 & 98.90 & 99.50 \\
WavMark    & 100.00 & 100.00 & 100.00 \\
\bottomrule
\end{tabular}
\caption{Gray-box cross-data results (\%). Top: attack success rate (ASR) of the original watermark. Bottom: recovery accuracy (ACC) of the overwritten watermark.}
\label{tab:graybox_crossdata_combined}
\end{table}

\begin{figure}[!t]
\centering
\includegraphics[width=0.9\columnwidth]{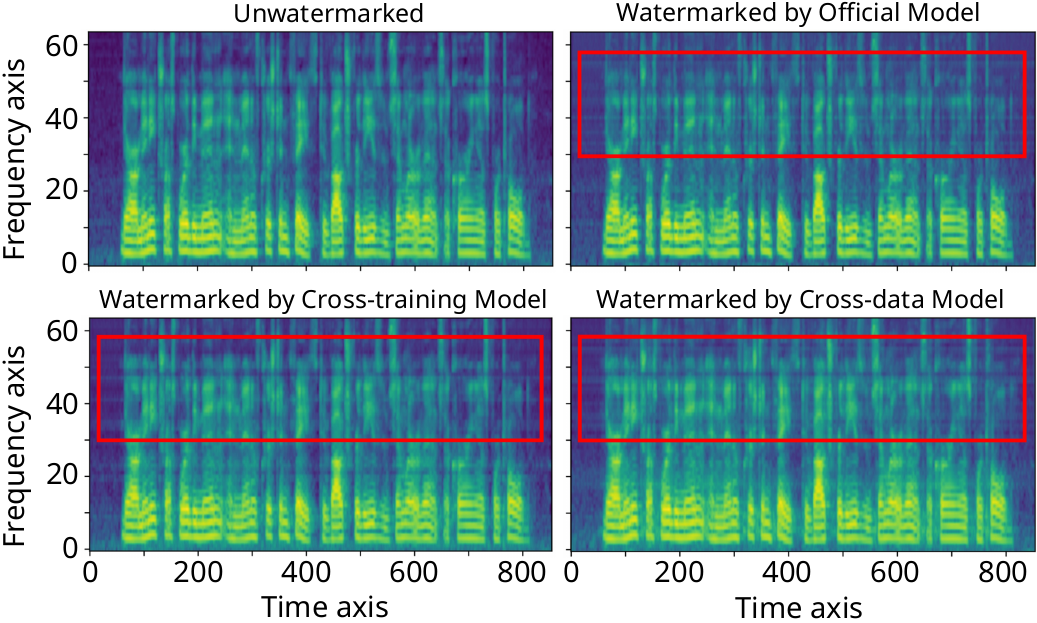}
\caption{Spectrogram comparison of the unwatermarked audio and three watermarked versions. Red boxes highlight the spectral perturbations introduced by the watermark.}
\label{spectrogram_comparison}
\end{figure}

\begin{figure*}[t]
\centering
\begin{subfigure}[b]{0.29\textwidth}
    \centering
    \includegraphics[width=\textwidth]{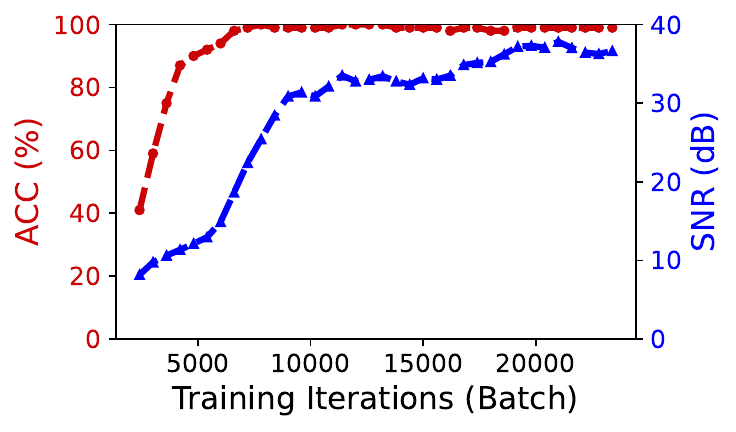}
    \caption{AudioSeal Reproduction}
    \label{blackbox_audioseal}
\end{subfigure}
\hfill
\begin{subfigure}[b]{0.29\textwidth}
    \centering
    \includegraphics[width=\textwidth]{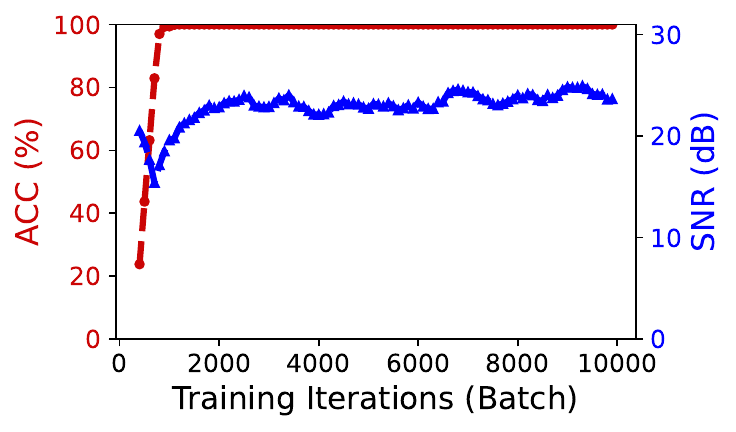}
    \caption{Timbre Reproduction}
    \label{blackbox_timbre}
\end{subfigure}
\hfill
\begin{subfigure}[b]{0.29\textwidth}
    \centering
    \includegraphics[width=\textwidth]{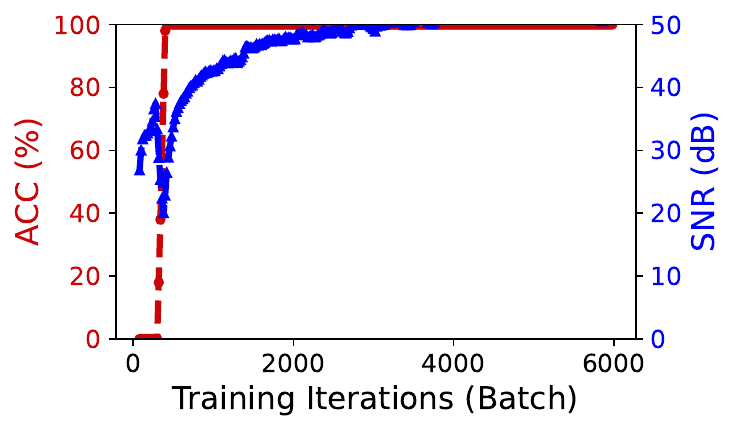}
    \caption{WavMark Reproduction}
    \label{blackbox_wavmark}
\end{subfigure}

\caption{Black-box reproduction attacks on three watermarking systems.}
\label{fig:blackbox_1x3}
\end{figure*}

In cross-data scenarios, Figures~\ref{graybox_crossdata_audioseal} and \ref{graybox_crossdata_timbre} illustrate that surrogate models trained on LibriSpeech effectively disrupt watermarks embedded by models trained on VoxCeleb1. The resulting BER distributions are centered around 0.5 for both AudioSeal and Timbre, indicating complete erasure of the original legitimate watermark.

Table~\ref{tab:graybox_crossdata_combined} quantitatively supports these observations, showing a near-perfect ASR and watermark recovery accuracy in all methods. The results demonstrate that surrogate models trained on varying datasets can effectively overwrite and replace original legitimate watermarks.

We further illustrate the spectrograms of unwatermarked and watermarked audio versions produced by official, cross-training, and cross-data Timbre models in Figure~\ref{spectrogram_comparison}. Despite differences in training details or datasets, all models embed watermarks within similar spectral regions (highlighted in red boxes). This consistent embedding behavior facilitates successful watermark overwriting, challenging the security assumption that secret training details or private weights are sufficient to ensure the watermark security.

\paragraph{Black-box Attack.}
In the black-box scenario, the adversary has no prior knowledge of the watermarking algorithm, model weights, or training data. We analyze two practical strategies: zero-query and query-based attacks. 

\begin{figure}[t]
\centering
\includegraphics[width=0.7\columnwidth]{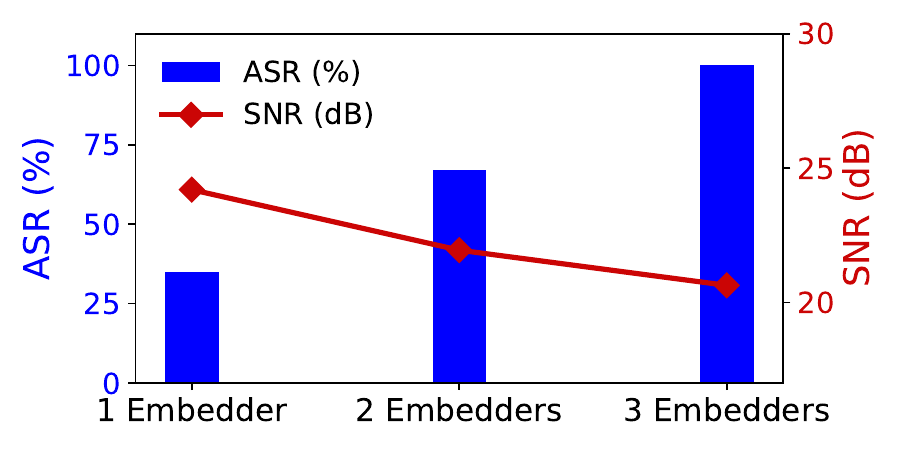}
\caption{Black-box attack success rate and audio quality.}
\label{blackbox_hist}
\end{figure}

Under the zero-query attack, the adversary collects or reproduces a set of candidate watermarking models and applies them sequentially in a brute-force manner to overwrite the original watermark. Figure~\ref{blackbox_hist} illustrates the impact of progressively stacking multiple watermarking methods, assuming a scenario with three common watermarking techniques. As more watermarking models are sequentially applied, the overwriting ASR increases from nearly 30\% (one embedder) to almost 100\% (three embedders), while SNR decreases from about 24\,dB to 20\,dB. Practically, as the candidate methods expand, the zero-query strategy becomes increasingly inefficient, leading to significant perceptual degradation and high computational overhead.

The query-based attack mitigates these drawbacks by utilizing limited queries to the watermark detector. Instead of exhaustively applying all candidates, the adversary iteratively ``try-and-test'' candidates and stops once the detector confirms successful overwriting of the original legitimate watermark. Consequently, audio degradation is limited to a single embedding operation. As depicted in Figure~\ref{fig:blackbox_1x3}, watermarking models exhibit varying convergence time. Certain methods reach sufficient overwriting capabilities early in the training, allowing effective overwriting with partially trained embedders. Table~\ref{tab:blackbox_attack_comparison} provides a comparative analysis between zero-query and query-based attacks. The results indicate that query-based attacks, using fewer than 10 detector queries, achieve a reduction of over 50\% in training iterations compared to the zero-query approach. By applying only a single effective watermarking method instead of stacking multiple methods, query-based attacks preserve the audio quality and maintain identical attack success rates. As the size of the candidate set grows, the advantages of the query-based approach become more pronounced.

\begin{table}[t]
\setlength{\tabcolsep}{4pt}
\centering
\small
\begin{tabular}{l|c|c|c|cc}
\toprule
\textbf{Attack Type} & \textbf{Query} & \textbf{Training Cost} & \textbf{SNR (dB)} & \textbf{ASR (\%)} \\
\midrule
Zero-query  & 0     & 36,000 iters   & 20.63 & 100 \\
Query-based & $<$10 & 14,000 iters   & 24.19 & 100 \\
\bottomrule
\end{tabular}
\caption{\small Comparison of zero-query and query-based black-box overwriting attacks.}
\label{tab:blackbox_attack_comparison}
\end{table}

\section{Discussion and Conclusion}
This paper presents the first systematic study of overwriting attacks, uncovering a critical security vulnerability in existing neural audio watermarking paradigms. Our results demonstrate that watermarking methods, while effectively optimized for robustness and imperceptibility, are vulnerable to intentional overwriting attacks. Consequently, audio ownership can be readily hijacked, significantly undermining watermarking's reliability for provenance verification.

The effectiveness of white-box attacks highlights a critical oversight: existing neural audio watermarking methods rely on model secrecy for security, neglecting explicit defenses against intentional manipulation. Our gray-box experiments further reveal that surrogate models, even trained on different data and implementation details, can consistently converge to similar embedding strategies. This architectural convergence undermines the assumption that the secrecy of model weights alone suffices for security. Additionally, our black-box evaluation confirms the practicality and feasibility of overwriting attacks under minimal knowledge and limited query access.

In conclusion, these insights highlight the necessity for a fundamental shift in neural watermarking research. Future neural audio watermarking methods must incorporate explicit mechanisms for security and integrate robust defenses against intentional adversarial attacks alongside imperceptibility and robustness goals.

\section{Acknowledgements}
 The work of L. Yao, C. Huang, and M. Pan was supported in part by the US National Science Foundation under grants CNS-2107057, CNS-2318664, CSR-2403249, and CNS-2431596. The work of P. Lin was supported in part by the National Science and Technology Council (NSTC) of Taiwan under grants NSTC 114- 2221-E-002-141-MY3, NSTC 112-2221-E-002-163-MY3, and NSTC 113-2314-B-A49-034-MY3. The work of T. Ohtsuki was supported in part by JST ASPIRE Grant Number JPMJAP2326, Japan.

\bibliography{aaai2026}
\end{document}